# A systematic literature review to unveil **u**sers' objective reaction to virtual e**X**periences: Complemented with a conceptual model (QoUX in VE)

**Alireza Mortezapour**

Department of Computer Science, University of Salerno, Fisciano (SA), Italy

**Andrea Antonio Cantone**

Department of Computer Science, University of Salerno, Fisciano (SA), Italy

**Monica Maria Lucia Sebillo**

Department of Computer Science, University of Salerno, Fisciano (SA), Italy

**Giuliana Vitiello**[1]

Department of Computer Science, University of Salerno, Fisciano (SA), Italy

---

[1] Corresponding author: Giuliana Vitiello, Email: gvitiello@unisa.it

## Abstract

In pursuit of documenting users' Neurophysiological responses during experiencing virtual environments (VE), this systematic review presents a novel conceptual model of UX in VE. Searching across seven databases yielded to 1,743 articles. Rigorous screenings, included only 66 articles. Notably, UX in VE lacks a consensus definition. Obviously, this UX has many unique sub-dimensions that are not mentioned in other products. The presented conceptual model contains 26 subdimensions which mostly not supported in previous subjective tools/questionnaires. While EEG and ECG were common, brain ultrasound, employed in one study, highlights the need for using neurophysiological assessments to comprehensively grasp immersive UX intricacies.

# 1. Introduction

It has been several years since people's interaction with technology has expanded beyond personal computers, laptops, and smartphones. For instance, the transition from two-dimensional to three-dimensional environments has occurred relatively recently [1]. These advances have occurred in less than 20 years and, to be optimistic, many of these developments have seen significant growth in the last 5 years [2]. The term *Metaverse* was first introduced in the science fiction novel Snow Crash in 1992, combining the words "meta" and "universe," and it appears to be gaining traction today [3]. Various researchers from different disciplines offer their interpretations of this concept. However, from the perspective of Human-Computer Interaction, it can be understood as a straightforward transition from a two-dimensional interaction to a three-dimensional interaction and immersive experiences for end users [4].

Experiencing an immersive 3D environment is no longer an unattainable dream for users; instead, it is increasingly becoming a reality for technology enthusiasts and members of the Z Generation [5, 6]. It can even be argued that no other year has been as remarkable as the present year, 2023. This year has witnessed significant advancements in technologies closely aligned with these concepts, bringing them closer to actualization. This year marked the introduction of Apple's Vision PRO glasses and similar offerings, such as the Meta brand's Meta Quest 3, thus providing consumers with tangible options. In addition to these hardware innovations, a multitude of notable software innovations also played a crucial role in satisfying people's desire to immerse themselves in the Metaverse.

As evident from the chronological progression of technologies related to the metaverse, the initial steps can be attributed to the advent of Virtual Reality (VR) mediums and VR glasses [7]. These technologies allow users to immerse themselves in a stable online 3D hybrid space, where they can interact with others as avatars [8]. Envisioning a future where millions, or even billions, of individuals, communicate, work, shop, attend school or university, and carry out various tasks within a virtual environment is not far-fetched [9]. The immersion within this emerging medium becomes more realistic when supported by some scientific evidence. Psychologists, for instance, have employed these technologies and reported significant positive effects on individuals [10, 11]. Moreover, researchers have claimed that their studies have proven highly beneficial in enhancing people's mood and productivity across various domains, including education and the workplace. These findings lend further credibility to the potential of these experiences and their impact on individuals' well-being and performance.

However, we, like other human-technology interaction specialists, firmly believe that a comprehensive understanding of any technology requires emphasizing the user [12]. Many experts widely hold the belief that the term User eXperience (UX) arguably encompasses the most comprehensive perspective regarding a user's interaction with technology [13, 14]. However, it is worth noting that some experts argue that the term UX has evolved from the concept of usability [15]. Since the initial mention of the term usability in 1979, there has been a wealth of published articles exploring the UX of various technologies, including 3D mediums [16]. Glimpsing Scopus with the combination of these terms, UX and Metaverse, leads you to the belief that more than 400 articles have been published. Or even if you want to search Google Scholar by only the Title of the articles, it increased to more than 600. Therefore, we can confidently assert that there is a growing emphasis on and swift progress in considering the UX within this emerging technology, metaverse and Virtual Reality [17].

The first place in this article in which we want to raise an important distinction is about a clear difference between two words: UX and usability. It is evident that the terms usability and UX are frequently used interchangeably, and sometimes even considered equivalent by researchers outside the realm of human-technology interaction [18]. However, experts in the field of Human-Computer Interaction are well aware of the inherent differences between these terms. Usability, as per the widely accepted international standard, encompasses aspects such as efficiency, effectiveness, and user satisfaction, focusing on the inherent characteristics of a system [19, 20]. In contrast, UX incorporates subjective opinions and psychophysiological responses of individuals regarding their interaction with a system or product [21]. Although it is recognized that both concepts must be considered collectively to ensure user-centered design, particularly in the context of ergonomic design, (in this study, of VR products), the present study highlights this distinction.

Our primary objective is to specifically review the UX of VR environments rather than encompassing the broader scope of usability analysis. Previous studies have neglected to acknowledge this distinction, compelling us to incorporate all the concepts related to both usability and User eXperience in our search strategy. We have thoroughly detailed this approach to compensate for the oversight and ensure comprehensive coverage of relevant literature.

One of the crucial aspects when evaluating the UX of a technology lies in the various methods employed to measure and quantify it. This matter holds such significance that a substantial portion of articles in this field typically delve into these diverse methods. These methodologies encompass a range of approaches, such as questionnaires, physiological measurements, cognitive assessments, and even evaluations of emotional variables [17]. And without a doubt, the technologies in the family of VR experiences also follow this principle. Numerous researchers have dedicated their efforts to exploring the UX of diverse users within VR mediums. Aligning with our assumptions in previous sentences, these researchers have employed a wide array of subjective and objective methods to quantitatively assess and measure this experience. Notwithstanding the multitude of studies that have underscored the correlation and affirmative correlation between subjective and objective outcomes, it is crucial to acknowledge that scholarly articles have been published elucidating the distinction inherent in these methodologies [22]. Hence, notwithstanding the convenience associated with gathering questionnaire and interview data, cultivating a comprehensive understanding of the physiological parameters influential in formulating an apt interaction with a VR system proves to be valuable.

Based on the most current information available, no published study has provided a comprehensive overview of the various psychophysiological methods, generally classified as objective methods, for investigating the UX of individuals within VR environments. While we recognize that any possible combination of our three family members ("user experience" AND "virtual reality" AND "objective assessment") may exist in the literature, it is important to note that in the next section we also reviewed these studies.

We formulated the following Research Questions (RQs):

***Research Question 1 (RQ1):*** *Which physiological signals are used in assessing the User eXperience of VR-based interactions?*

The investigation focuses on identifying the physiological signals utilized in the assessment of User eXperience in VR-based interactions.

***Research Question 2 (RQ2):*** *Which neurological signals are used in assessing the User eXperience of VR-based interactions?*

The aim is to determine the neurological signals employed for evaluating the User eXperience of VR-based interactions.

## 2. State of the art

In this section, we reviewed the existing literature, including previous systematic reviews.

### 2.1 UX in Virtual Reality

As extensively explained in the methodology section, in accordance with the COSMIN guidelines, we have integrated concepts within the scope of usability into our research strategy. As a result, a considerable number of articles in this area focus primarily on evaluating the effectiveness and efficiency (as is well known, these two elements are subsets of usability) of VR systems, especially in the context of medical interventions [23, 24], which are beyond the scope of this particular section of the article. However, some noteworthy review articles that have relevance to our study are presented below. However, as mentioned above, UX in VR-based interactions has not been extensively explored in existing studies.

2.1.1 Usability evaluation of health-related applications using Virtual Reality has garnered significant attention in [25]. The authors highlighted that evaluation techniques to evaluate the usability of Virtual Reality -based health interventions are significantly lacking compared to other digital health technologies. Although their specific search strategy remains unclear, they referenced a minimum of 25 key articles in their study. Drawing on insights gleaned from these reviewed articles and existing general guidelines on usability evaluation, the researchers proposed six distinct approaches that are suitable for evaluating usability in Virtual Reality-based healthcare technologies. These approaches include cognitive or task procedures, graphical evaluation, post hoc questionnaires or interviews, physical performance evaluation, user interface evaluation, and heuristic evaluation.

Furthermore, they strongly recommend the inclusion of specific user groups, such as external users who are not directly involved in the development process, in some evaluation methods. They also highlight the importance of engaging representative users who accurately represent the intended end-user population.

2.1.2  In a recent study which is published in 2022, three Brazilian researchers explored the benefits and challenges of virtual-reality-based usability testing and design reviews in industry through a patents and articles review [26]. In the final stage, they reviewed 7 patents and 20 articles. They implied that the industry which has used these ones, is automotive industry. Also, they declared that HTC VIVE head-mounted device which is frequently paired with motion capture systems and Unity 3D game engine is the most frequent used device for usability testing. VR benefits usability testing by providing the visualization of new angles that stimulate novel insights, increasing team engagement, offering more intuitive interactions for non-CAD specialists, saving redesign cost and time, and increasing participants' safety. Also, the main challenge of VR scenarios which are discussed in this article are a lack of realism due to unnatural tactile and visual interactions, latency and registration issues, communication difficulties between teams, and unpleasant symptoms. In this study 3 main usability testing are referred based on their frequency in the past articles including "time to complete the task", "system usability scale (SUS)", and "number of mistakes" of the users when did a scenario.

In a recent study published in 2022 [26], the authors explored the benefits and challenges of VR-based usability testing and design reviews in the industry through a review of patents and articles. In the final phase they examined 7 patents and 20 articles, suggesting that the industry that used them was the automotive industry. Additionally, they stated that the HTC VIVE head-mounted device, often paired with motion capture systems and the Unity 3D game engine, is the most frequently used device for usability testing. VR benefits usability testing by viewing new angles that stimulate new insights, increasing team engagement, offering more intuitive interactions for non-CAD specialists, saving redesign costs and time, and increasing participant safety. Furthermore, the main challenges of the VR scenarios discussed in this article are the lack of realism due to unnatural tactile and visual interactions, latency and registration issues, communication difficulties between teams, and unpleasant symptoms. In this study, 3 major usability tests are mentioned based on their frequency in previous articles, including "time to complete task", "system usability scale (SUS)", and "number of errors" of users when perform a scenario.

## 2.2 Psychophysiological response of body in Virtual Reality

Undoubtedly, immersion in Virtual Reality environments and experiencing a 3D space have a profound impact on individuals, eliciting a wide range of emotional, physiological, and cognitive effects. Extensive research has been conducted to explore the various implications of engaging with these simulated environments on the human body. While some studies have highlighted negative effects such as motion sickness, headache and nausea, others have revealed positive cognitive and physiological outcomes, particularly in therapeutic applications. While a detailed examination of these effects is beyond the scope of this article, it is worth noting that several studies have used psychophysiological methods to capture and analyze User eXperiences in VR. Functional magnetic resonance imaging (fMRI) has been used to observe brain activity during VR experiences, while other neurological techniques such as electroencephalography (EEG) and functional near-infrared spectroscopy (fNIRS) have also been employed. Additionally, researchers recognized the importance of evaluating physiological responses, including heart rate and heart rate variability, as indicators of the body's reaction to VR stimuli. Based on these considerations, it is logical to use neurophysiological methods to investigate the main psychophysiological response of the body when evaluating the usability and User eXperiences of VR means. Therefore, this study aims to use this approach to answer the main research question: "What is the body's primary psychophysiological response in evaluating UX in Virtual Reality environments?"

## 2.3 Psychophysiological measures of UX

2.3.1 Limited to the period of March 2020, a review article has been published regarding neurological and physiological measures of UX in the context of "Information System" [18]. The main difference between this article and the present one can be seen in the difference in context. These authors studied the usability and UX of information systems through psychophysiological variables, which is far from the scope of our study. They started with 957 articles and concluded with 27 articles for full-text review. The results indicate that UX assessment based on neurological and/or physiological measures has gained significant attention in the field of information systems. Over the years, the results of these studies have increasingly underscored the importance of evaluation-oriented guidelines in this area. The main findings of this study with respect to its objective are:

- Brain wave/EEG analysis is recognized as the main physiological measure of usability.
- As for neurological measures, they include facial expression, emotion, facial coding, eye tracking, galvanic skin response, electrodermal activity, heart rate variability, ECG, electromyography, and muscle activity.
- Except for muscle data, which was significantly less used, the rest of the measurement methods were used approximately equally in the included studies.
- The authors devoted a section to presenting important devices from different companies used for measuring cardiac activity.

2.3.2 In [27], the authors published a review on the application of EEG and fMRI to measure aesthetic processing in the user's visual experience. Since their context is related to 2D human-technology interaction, they considered the field of visual interaction in terms of art therapy, information visualization, websites or mobile applications, and other interactive platforms. They believe that their findings are useful for designers, artists and engineers using brain-computer interface art technology in the visual interaction experience. They discussed the two main models of visual aesthetics: "Chatterjee's model of the neural basis of visual aesthetics" and the "model of aesthetic appreciation and aesthetic judgment." Examining the brain, they revealed that all four main lobes of the brain, including Frontal, Parietal, Temporal and Occipital, are important in UX/UI evaluations. They further stated that each region of the lobes plays a different role in aesthetic appreciation. Furthermore, they stated the role of different parts of each lobe in UX/UI studies, which is beyond the scope of this part of the article. Furthermore, when they presented information on the application of EEG, they implied some important findings which are summarized here:

- Better UX evokes stronger relative power of delta in frontal lobe
- Better UX affects weaker relative power on the theta rhythms frontal and parieto-occipital regions.
- The beta frequency band is associated with emotions in the immersive or virtual environments.
- Better UX evokes stronger relative gamma power in C3 region of the brain.
- Better UX evokes stronger relative alpha power in frontocentral, parietal, and parieto-occipital parts.
- Also, the authors presented detailed information for the main regions of brain which are responsible for each of UX components including *initial aesthetic appreciation, delayed aesthetic appreciation, etc.*

2.3.3 In another recently published review that focused on neuroimaging techniques, aesthetic appreciation garnered attention [28]. The authors shared the belief that aesthetic appreciation is not an isolated neurobiological process exclusively devoted to evaluating particular objects. Instead, they put forth the notion that it represents a broader system, primarily centered on the mesolimbic reward circuit, which plays a role in assessing the hedonic value of diverse sensory objects. Furthermore, the authors declared that neuroscientific research highlights that hedonic values are not solely determined by the properties of objects themselves but are also influenced by various external factors that modulate the subjective experience.

As a concluding mark, we can discuss that, as we progress from the first study, which is the most closely related, to the third study, we observe a decrease in the semantic proximity to our own study. Hence, it can be deduced that despite the existence of valuable published studies, none of them have provided a definitive answer to the main question addressed in our present study. This further affirms the novelty and innovation of this study.

## 3. Methodology

Conducting a systematic review involves adopting an inclusive approach to the literature related to a specific topic, while simultaneously adhering to the highest standards of rigor. In this article, we analyze published articles in the field of psychophysiological response of the body for the evaluation of UX in the virtual world. For avoidance of unnecessary repetitive publish, we checked three major systematic review registry databases, including PROSPERO, the Register of Systematic Reviews/Meta-Analyses in Research Registry, and INPLASY. This step ensured the team about novelty of this research.

In this systematic review, the COSMIN methodology for systematic reviews of patient-reported outcome measures and the Preferred Reporting Items for Systematic Reviews and Meta-Analyses (PRISMA) were used as methodological guides. The main one was PRISMA but in some cases we used the COSMIN guidelines to ensure the completeness of our study, for example by using a larger number of keywords to reduce the possibility of missing articles mentioning UX, especially through similar concepts such as usability. The PRISMA checklist completed for the current study is presented in Appendix 1.

This study was conducted in six steps:

1- Formulating the main question.

2- Development of search strategy.

3- Search the databases and extracted the final articles.

4- Initial monitoring of articles based on the eligibility criteria AND Title/Abstract.

5- Quality check of the remaining articles / read all of them in full text.

6- Synthesis of the results.

### 3.1 Formulating of the main question

As mentioned, at the end of this study, we want to find out what physiological responses the human body shows in response to evaluating the UX of VR environments. Therefore, to answer this question, we must examine articles from 3 separate families of keywords in combination with each other. Physiological responses of the human body are in the first family, UX is in the second family, and Virtual Reality environments are in the third family.

### 3.2 Development of the search strategy

In this study, our sole objective was to focus on the User eXperience. However, during the initial research phase, we recognized the need to incorporate related concepts such as usability, including its subsets, to ensure a comprehensive search strategy. While we are aware of this inherent difference, we have deliberately included subsets of usability such as efficiency, effectiveness, and user satisfaction to minimize the possibility of overlooking relevant articles. This decision, while challenging for the document review team, was crucial to achieving a thorough and complete review.

As is evident, we have three distinct families of keywords that should be combined with respect to Boolean logic. According to the guidelines of the COSMIN methodology for systematic reviews, subscales of the main keywords can also be assessed. Therefore, we use some subdimensions of usability in the search strategy, including effectiveness, efficiency, and user satisfaction. The current research team knew that this step would retrieve many articles through the false positive mechanism. But to have a complete list of published articles, we accepted the risk of using wider keywords.

Therefore, we first conducted an initial search with the main keywords of each family, then, with a brief review of the most important published articles, we also extracted the relevant sub-keywords. Then, by checking the synonyms of all these keywords (main and secondary), we wrote the initial search strategy. Ultimately, this search strategy was reviewed with several industry experts. 18 keywords in the "usability and user experience" family, 29 keywords in the "physiological body responses" family and 8 keywords in the "virtual reality environments" family were used. We did not use the acronym "VR" as a search term on the assumption that it usually cannot appear in the title and abstract. However, we tried to complement our keywords with other family members to ensure the completeness of our search. After checking the first results, we believed that only the acronym for Virtual Reality (VR) could produce false positive results. We have therefore included other acronyms such as fMRI, fNIRS etc. in our research strategy.

It is obvious that in some databases some of these keywords could not be used, in which case their use was omitted. To see the search strategy used in this study, see Appendix 2.

### 3.3 Selecting the appropriate databases

Considering that we focused on 3 main word families in this study, numerous review articles from each of these families or in pairs were checked to ensure the main databases needed to initiate the search. Accordingly, systematic searches were carried out in seven databases, as shown in Table 1. To complete the search, a manual search was also conducted on Google Scholar and the bibliography list of the included primary studies.

The choice of databases and journals was guided by the multidisciplinary nature of the research focus, so we chose these databases because they contain a large amount of scientific literature spanning the fields of neuroscience, psychology, human-computer interaction and engineering/technology. Web of Science and Scopus were the two large bibliometric databases with broad coverage of subject areas used by all authors. Considering that we also had to search for medical sources, Medline was also added to this list. ACM and IEEE cover subject areas in engineering disciplines and can be useful for our goal of comprehensiveness. Based on our initial research to produce our initial list of keywords, we encountered some articles from underutilized databases such as ProQuest and Ebsco. We therefore believed that searching these databases could be useful for our work.

*Table 1. Complementary information about databases*

| Num. | Name of the Database | Search through | Initial retrieved articles |
|------|---------------------|----------------|----------------------------|
| 1 | Scopus | Title-Abs-Key | 1447 |
| 2 | PubMed | Title-Abs | 364 |
| 3 | Ebsco | Abstract | 232 |
| 4 | Web of Science | Topic | 824 |
| 5 | IEEE explore | All metadata | 321 |
| 6 | ACM digital library | Abstract | 33 |
| 7 | ProQuest | Abstract | 181 |
| 8 | Google Scholar | - | 10 |

## 3.4 Monitoring the remained articles

The collection of articles in the previous phase ended with 3412 articles. Of these, 1,592 articles were repeatedly indexed. These articles were initially removed from Endnote, using the Find Duplicates function, and then manually. Subsequently, another 93 articles were excluded because this was a review, and finally, 1727 articles were initially reviewed for this phase. Without eliminating another article, we decided for each of them based on the title/abstract. If they were eligible, we considered them for the full-text review phase and did not eliminate them. Our main criteria were:

1) Articles that aimed to consider the psychophysiological response of the body to measure the UX of the VR-related interaction (or user satisfaction subset of the usability construct that can be a sign of an experience positive user),

2) Published in English,

3) Using psychophysiological method when using VR scenarios. We specified during use why they needed to measure the efficiency and effectiveness of these interventions; for this reason, since many studies used physiological data before/after the use of VR-based interventions, they were excluded.

To make the articles included in this study clearer, as well as to complete the inclusion criteria, we also considered a series of exclusion criteria, which include:

1) Not using psychophysiological measures for the evaluation of UX, for example using only a related questionnaire,

2) Use psychophysiological measures only for the evaluation of the efficacy and effectiveness of a VR scenario and therefore do not report the UX or even the user satisfaction subset of the usability

Figure 1 shows the Prisma flowchart.

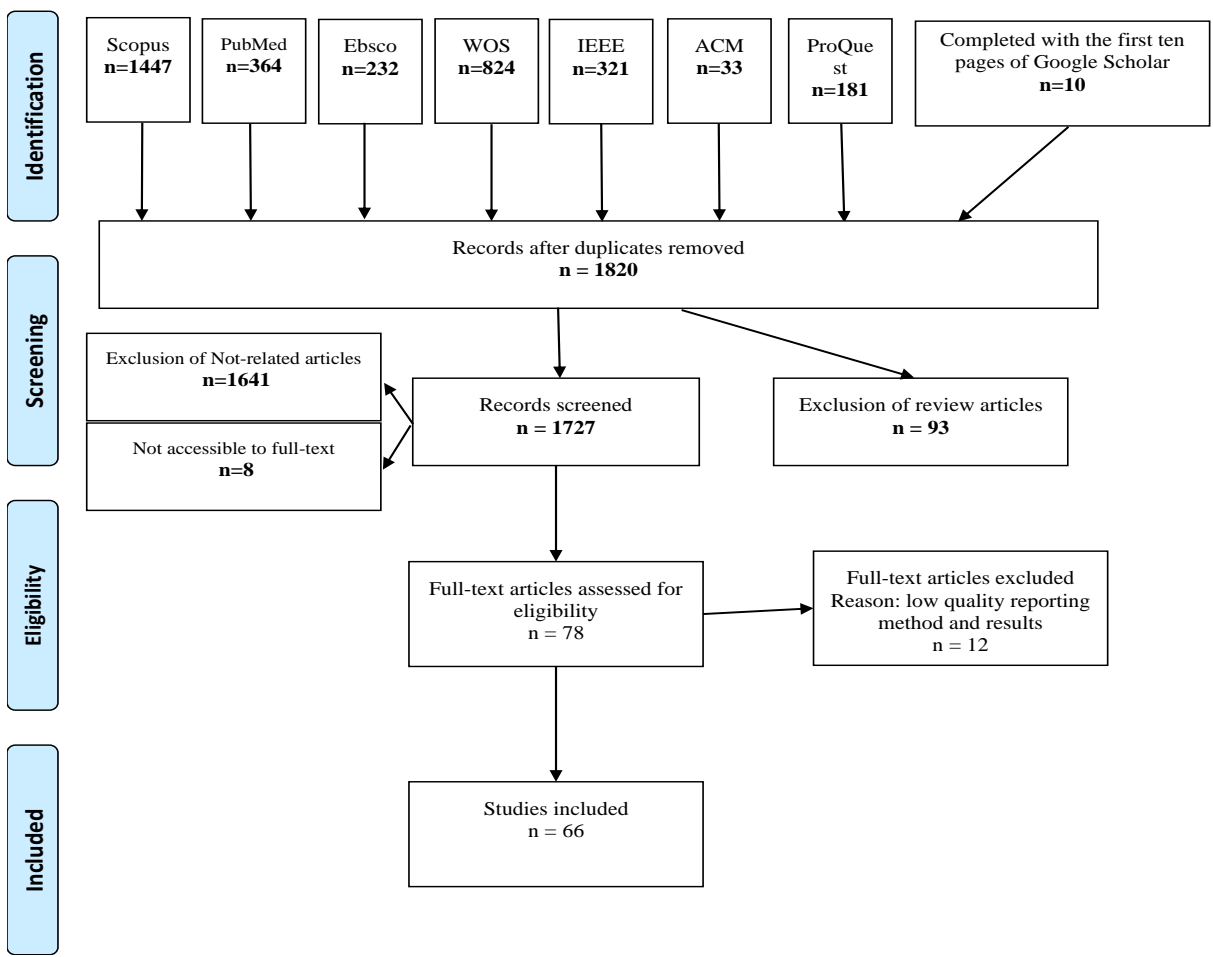

*Figure 1.PRISMA flowchart*

### 3.5 Full text reading of remained articles

After initial assessment of articles by title/abstract by two independent authors to reduce bias, 78 articles remained for full-text review. We expected that if there were different opinions on an article, disagreements would be resolved by consensus, but we did not find any inconsistencies. There was 100% agreement on the inclusion of articles between the two reviewers.

It is evident that these articles are directly related to the evaluation of the usability of virtual environments and the experience of end users when interacting with these mediums based on the neurophysiological response of their body.

### 3.6 Synthesis of the results

Based on evaluations, only 66 articles remained for the final evaluation step. Details from the included publications were extracted and collected in a spreadsheet, and categorized into three sections: general information, methodological details, and results.

## 4. Results

After double-checking the remaining articles, the information on these 66 articles is summarized in the following sections. During the full text review of these articles, by checking their references, we could not find any other suitable articles to add to the previous 66. This method, called reference checking, is an acceptable way to increase the chances of including fully relevant articles [29, 30].

Considering all these articles, overall, 26 different experiences with virtual environments (both Virtual Reality and Augmented Reality) were found. These 26 variables, according to the authors' claims, define different user experiences in the end consumer of these products. For most of these usage experiences, different measurement methods based on the physiological data of the end user were presented. However, for some of them, questionnaire data are the only method of quantification.

### 4.1 Metadata information

The oldest remaining article dates back to 2002, and 7 new articles were published in 2023, indicating an increase in this interesting topic.

In total, 323 authors published these 66 articles collaboratively within their research group. Figure 2 shows the analysis of co-authorship of these researchers in relation to the year of publication. As can be seen, as we approach the last few years, the number of co-authors in an article has decreased.

Figure 3 illustrates the co-occurrences of keywords in these articles. The total number of keywords submitted by the authors of these articles is 220. As shown below, 146 of them have significant co-occurrence with each other. The EEG was the most widely used device to study different user experiences within virtual environments. As evidenced by the light green color, the application of new methods such as fNIRS and EDA/GSR has increased in recent years. On the other hand, immersion and presence are the most cited topics in related literature.

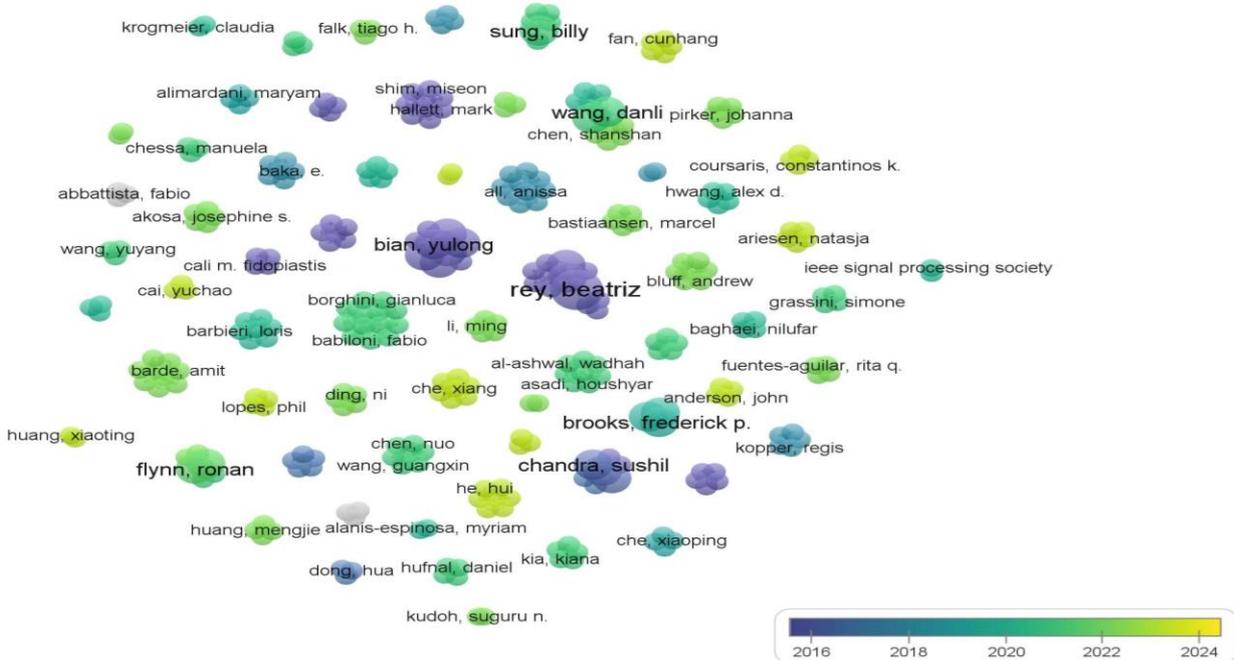

*Figure 2. Co-authorship analysis of remained articles*

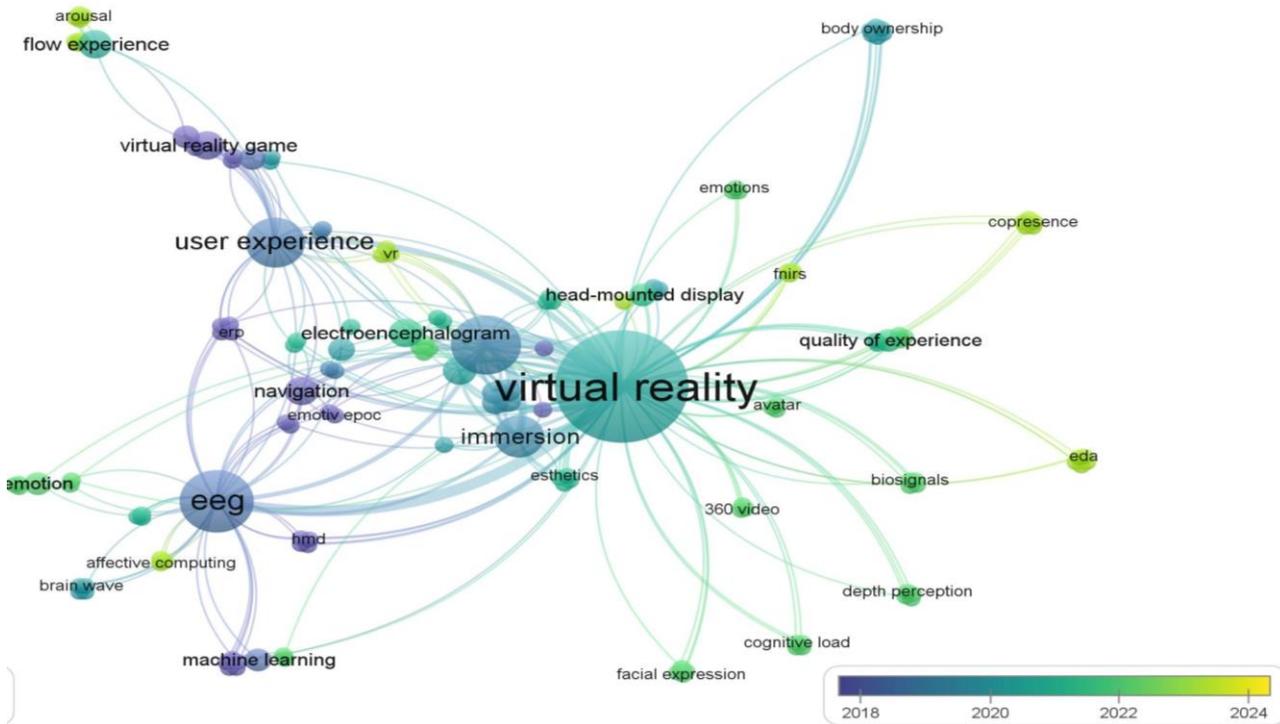

*Figure 3. Co-occurrence of keywords*

## 4.2 Methodological points

The number of participants in the included studies was generally less than 100; only 3 studies had a higher number of participants. A total of 2218 people participated in these 66 studies. Although the gender of 270 of them was not stated, participants included 846 women and 1102 men. The mean and median of the participants were 33 and 27 years old, respectively. These values are important for the design of further studies. The smallest sample size was for a 2007 study with 2 participants (1 M and 1 F) and the largest was for a 2021 study with 248 participants.

The minority of included studies, mostly old, use devices such as caves, walls and screens to produce virtual experiences for participants. In addition to these, most studies use new immersive devices such as goggles and masks. In the latter category, most researchers use HTC Vive, HTC Vive Pro (eye) and Oculus Rift (over 70 percent). Other devices such as Meta Quest 2, Sony PlayStation and nVisor SX were used in only a few studies.

In the included studies, some devices for measuring users' body physiological parameters were used only once, such as brain sonography, or only twice, such as fMRI. However, some devices such as EEG, ECG and EDA/GSR were used more than 10 times each. More details about these devices and their characteristics are given in the following sections.

## 4.3 The main points

Although precise definitions of UX have been provided in several products, in products based on virtual environments, including Virtual Reality and Augmented Reality, there is still no consensus to define UX and its various aspects. Meanwhile, there is agreement on some different aspects of the UX of these environments, among which we can mention Immersion, Cybersickness and Flow. However, there is still no consensus on some aspects such as the mental load of the user. But what became apparent to us is that some aspects of the UX that may not have existed in other products before were crucial sub-dimensions of the UX here.

In general, we found that 26 different categories can be considered as UXs of virtual environments, presented in Table 2 and complemented by their respective definitions. In the next part, we presented neurophysiological evidence for these UX parameters.

**Table 2. Various sub-dimension of UX in virtual environments**

| Number | UX sub-dimension | Definition applicable to virtual environment | Repetition in included studies |
|--------|------------------|---------------------------------------------|--------------------------------|
| 1 | Immersion | perception of being physically present in a non-physical world. | 4 |
| 2 | Presence | The ability of an end-user to feel that they are actually in a virtual location and experiencing a non-real event. So, it consists of two types of illusions name place illusion and plausibility illusion. | 19 |
| 3 | Embodiment | possibility in VR to visually substitute a person's real body by a virtual one, seen from the person's own first-person perspective. | 1 |
| 4 | Engagement | Engagement is a multidimensional construct that refers to an end-user's active involvement in a virtual environment. | 8 |

| 5 | Coherence (coherent interaction) | End-user's perception of the level of harmony in the virtual environment. A VR system with a high level of coherence could lead to great illusion. | 2 |
|---|---|---|---|
| 6 | Flow | Optimal user experience (highly positive mental state) of effortless attention and enjoyment when immersed in a virtual environment. | 3 |
| 7 | Cognitive load | The amount of mental effort of the user from his/her mentally available resources while being in the virtual environment. | 4 |
| 8 | Sense of unity | A sense of integration between the reactions of people presents in a virtual environment (such as an online meeting or virtual concert). | 2 |
| 9 | Sense of ownership | End-user's experience about replacement of their own body with a virtual body inside a virtual environment. | 1 |
| 10 | Sense of oneness | It has equal definition with "sense of ownership". | 1 |
| 11 | Sense of agency | Refers to the feeling of controlling one's movements and, through them, the events in the virtual world. | 3 |
| 12 | Believability | The extent that a virtual environment can meet the expectations of its end user compared to the real world. But over time, users can ignore small system errors. | 1 |
| 13 | Difficulty experiences | Perceptual judgment about the level of hardships of a virtual content e.g., Virtual reality games. | 2 |
| 14 | Intuitive interaction | An interaction with the virtual environment is called an intuitive interaction where the components and content of that virtual environment have already been experienced by the user in the real environment. It means that the understanding of that content is tangible for the user. So, they can retrieve better how to perceive them. | 1 |
| 15 | Flexibility | The degree of perceived deviation of virtual content from end-user's real expectations. For example, to what extent a character can deviate from its actual behavior. | 1 |
| 16 | Emotional response | Conscious positive and/or negative response of end-user's body in response to virtual content. | 6 |
| 17 | Cybersickness | A set of discomforts perceived by the end user during or after using a virtual environment. It has many symptoms including nausea, dizziness, blurred vision, etc. This condition mainly results from the conflict between the person's expectations and the visual content presented to his/her in the virtual environment. | 5 |
| 18 | Perceived instability (Aliveness) | refers to all unrealistic sensations (such as the apparent aliveness of a surface) that cannot be attributed to the physical properties of a textured surface rendered with a force-feedback device. | 1 |
| 19 | Visual fatigue syndrome | Uncomfortable symptoms such as eyestrain, dizziness, and other visual problems with visual interaction with virtual environment. | 1 |
| 20 | Free navigational experiences | It is a negative type of information seeking experience that occurs when an end-users cannot retrieve suitable spatial information during presence in the virtual world. | 3 |
| 21 | Vergence-accommodation conflict | It is a negative experience that an end-user reports it when his/her brain receives mismatching cues between the distance of a virtual 3D object (vergence), and the focusing distance (accommodation) required for the eyes to focus on that object. | 2 |
| 22 | Enjoyment | It a positive satisfaction of a human being needs when interaction with a virtual world. | 1 |
| 23 | Arousal | Particular positive or negative feeling when interaction with a virtual environment. It occurs when a such product cause someone to have a particular feeling. | 7 |

| 24 | Affective interaction | It's a positive or negative consequence results from interacting with virtual environment. It depends on the emotion of the end-users when using the product. | 3 |
| 25 | Enrich interaction | It's a positive feeling that occurs when an end-user experiences a holistic multimodal interaction with a virtual world not only visually but also with tactile feedbacks. | 2 |
| 26 | Intimate interaction | It's a positive response of an end-user regards to an avatar in a virtual world. | 1 |

Regards the first two rows of Table 2, it is worth mentioning that sometimes the two terms immersion and presence are used interchangeably. Immersion indicates the objective level of sensory fidelity provided by a VR system, while presence is subjective [31]. Some researchers have claimed that presence is a consequence of immersion. In addition, others believe that immersion is a "technology-related" concept, while presence is a "human perception-related" concept. Here, we present a conceptual model for the proposition of user experiences in virtual environments (figure 4), Quality of user experiences in a virtual environment: **QoUX in VE**. However, it's evident that with the progress of studies and research around virtual environments, the number of UX sub-dimensions can be increased.

In addition, to measure these user experiences, various methods based on the physiological parameters of the users' bodies have been introduced, which are presented in the next section.

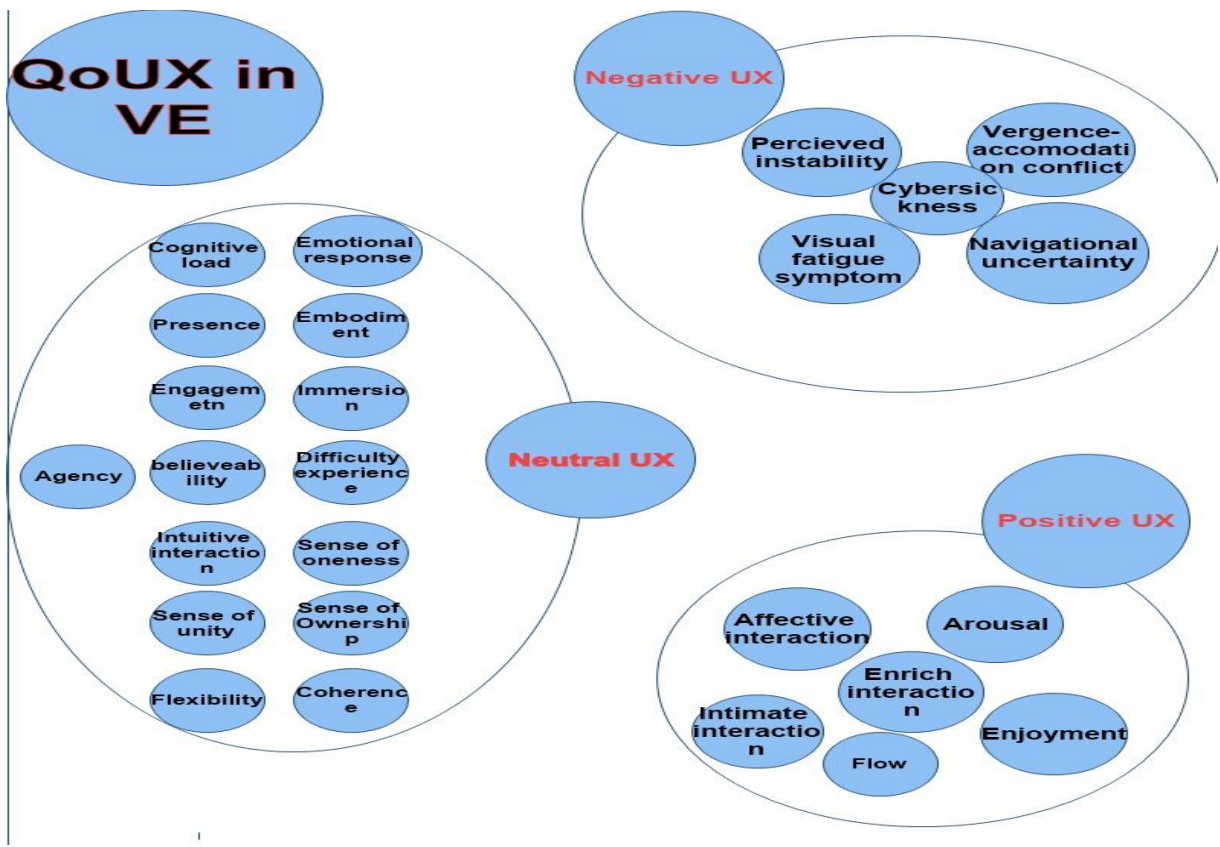

*Figure 4. QoUX in VE conceptual model*

As it is clear from the proposed model, the user's experience of a virtual environment consists not only of its positive parameters, contrary to the opinion of some authors, but also because of the limitations of the systems that provide 3D space, as well as the limitations of the user's body in adapting to these environments; for example, the user's experience has negative experiences such as cybersickness.

Also, in one part of the model, elements that are not positive or negative are placed in a neutral form, which can be positive or negative depending on the conditions.

## 4.4 Neurophysiological evidence

This section briefly presents the results of the included studies related to the evaluation of the UX of virtual environments and its subsets through users' physiological parameters (table 3).

**Table 3. A glimpse on main points in the included studies**

| No. | year | Methods and techniques | Participant | UX sub-category | Setting | Main results |
|---|---|---|---|---|---|---|
| 1 | 2023 [1] | EEG - Alpha Active Ltd device – 2 channels (AF7 and AF8 in Frontal lobe) – 128 Hz sampling rate – Analysing Alpha and theta/beta ratio. | 14 (10 M, 4 F) | Sense of presence | VR (Oculus Rift) | **Alpha** power: 5.12, 4.66 and 4.90 respectively for fearful, relaxing and exciting scenario. **Theta/beta** ratio: 3.66, 3.66 and 3.84 respectively for fearful, relaxing and exciting scenario. **The** results indicated strong links between presence and generation of fear, which is vital in the efficacy of exposure therapies. |
| 2 | 2023 [2] | HR and ResR- iMEC 12 device. | 170 (106 M, 64 F) | Arousal / Enjoyment | VR (not-declared) | **HR** and ResR are two important physiological measure in studying emotional response of users of the virtual tourism. |
| 3 | 2023 [3] | ECG - HR - 256 Hz sampling rate. EDA - Cobalt System device. | 28 (16 M, 12 F) | Arousal | VR (Oculus Quest 2) | **EDA** was higher during active than during passive VR for solo participants, whereas this significant effect was not detected in dyads. |
| 4 | 2023 [4] | EEG – 8 electrodes (F3, Fz, F2, C3, C4, P3, Pz, P4) – Analysing Beta and Gamma waves. BR (breathes per minutes) – By a trap on the chest. GSR (mS) – Imotions device. HR (beats per min). | 10 (6 M, 4F) | Immersion | VR (Oculus Quest 2) | **Based on** EEG data, adding physical probe (tactile interaction) did not have a significant impact on the participants' brain activity.<br><br>**Based on** respiration data, using a physical prop in a virtual environment enhances involvement and immersion in the virtual world.<br><br>**Based on** GSR data, the authors claimed that the amount of sweat of the participants can be a sign of UX in the VR.<br><br>**Based on** HR data, the authors claimed that HR can be a good index for assessing some aspects of UX in the VR.<br><br>**A physical** object does indeed add depth to the virtual experience. |
| 5 | 2023 [5] | ECG - HRV (R-R interval) – both time domain and frequency domain – SOMNO | 24 F | Flow experience | VR (HTC Vive Pro) | **Lower** low frequency (LF) and SDNN components of HRV indicated a flow experience, demonstrating the provision of more attentional resources. **Higher** flow experience results in higher learning efficiency (better performance) in Three-dimensional multiple object tracking. |

| | | | | | |
|---|---|---|---|---|---|
| | | touch™ RESP (SOMNO medics) - 512 Hz sampling rate. | | | | |
| 6 | 2023 [6] | EDA in microsiemens (µS) - HR – Both measured with Empatica E4 wristband. | 10 | En- gagement | VR (MetaQ uest2) | **EDA** variability for all participants throughout the VR experience 0.11–5.99 µS with baseline variability established as 0.24–0.97µS. **Phasic** changes (compared to tonic) in the EDA signal appearing as a direct response to stimulus in the VR. **Participants** show an increase in EDA signal, which signifies an enhanced level of excitement or engagement. |
| 7 | 2023 [7] | EEG - EPOC+ EMOTIV device- 14 channels – AF3, AF4, F3, F4, F7, F8, FC5, FC6, T7, T8, P7, P8, O1, O2 regions of the scalp - 128 Hz sampling rate. | 28 (7 M, 21 F) | Emotional response compromi sed from focus, engageme nt, relaxation and interest | AR (on a Huawei mobile device) | **ENG** index (ENG is characterized by increased physiological arousal and beta waves along with attenuated alpha waves) = 2.044 **MED** index (the ability to switch off and recover from intense concentration) = 1.006 **INT** index (the degree of attraction or aversion to the current stimuli) = 1.829 **FOC** index (fixed attention to one specific task) = 1.090. **Learning gain** (subjectively assessed) with the AR help, significantly correlates with ENG, INT and FOC. |
| 8 | 2022 [8] | ECG (HR and HRV) - LiveAmp biosignals amplifier (Brain Products GmbH) – 500 Hz sampling rate. | 8 | Sense of presence | VR (HTC Vive) | **Average** HR = 90.75 – Average HRV = 16.48. **Due to** the limited number of participants, it was not possible to make strong claims about the correlation between presence and HR or HRV. |
| 9 | 2022 [9] | EEG – 64 channel - 1000 Hz sampling rate. | 15 (7 M, 8 F) | Stereosco pic visual fatigue (SVF) | VR (HTC Vive Pro) | **Amplitude** of the P2 (distributed to partial lobe) which originated from posterior cingulate cortex and precuneus, having strong functional correlation with depth perception as well as with self-awareness was significantly associated with SVF. **SVF** is rather a conscious status concerning the changes of self-awareness or self-location awareness than the performance reduction of retinal image processing. |
| 10 | 2022 [10] | fNIRS – Assessing blood oxygenation variations in the brain - Biopac F2000. EDA/SCR, HR - Biopac MP-160 with an electrodermal amplifier included EDA100C. | 60 (12 M, 48 F) | Arousal (one of the main affective dimension s) | VR (Vive Pro-eye) | **fNIRS** can be good choice for studying cognitive load in the virtual environment. **The** heatmap of the attention generated by eye-tracker of the VR device can be a good index for studying attention in the virtual environment. **For** skin conductance responses (SCR) and heart rate (BPM) there were statistically significant **Difference** between the static and 360 VR images (360◦ VR images elicited stronger arousal). **Pupil** dilation results showed that 360◦ VR images generated greater pupil dilation in participants, than did the static VR images. **Considering** that many of the questionnaires used in this study were unable to determine the difference between 360 and static images, as a result, there is a need to use physiological variables of users to measure their user experience. |
| 11 | 2022 [11] | EEG - ANT Neuro equipment | 24 (14 M, 10 F) | Emotional response | VR (HTC Vive | **Compared** with negative emotions, the alpha band power in the frontal lobe of the brain, beta and gamma bands in the temporal lobe region is |

| | | device – 32-channel. | | | Pro) | significantly higher under positive emotions.<br>**EEG** is a good measure for studying the emotions of the leaners in the VR environment. |
|---|---|---|---|---|---|---|
| 12 | 2022 [12] | EEG - 32-channel. | 32 (16 M, 16 F) | Sense of agency (SoA) | AR (12.7-inch touch-based tablet with installed AR applicat ion) | **Correlations** between SoA and translation modes were explored on handheld AR interfaces, revealing that the 1DoF translation mode was associated with higher SoA.<br>**EEG** spectral power analysis was used to compute alpha, beta, and gamma power, which were connected to SoA evaluation.<br>**Significant** increases in alpha power throughout the whole brain from 1DoF to 3DoF modes.<br>**Significant** differences in beta power were observed in certain regions when participants employed 1DoF mode compared with 3DoF mode.<br>**Significant** increases in gamma power were mainly observed from 1DoF to 3DoF across the whole brain. |
| 13 | 2022 [13] | BVP - 64 Hz sampling rate.<br>IBI - 1Hz sampling rate<br>HR - 1 Hz sampling rate.<br>GSR/EDA - 4 Hz sampling rate.<br>XYZ raw acceleration - 32 Hz sampling rate.<br>ST - 4 Hz sampling rate<br>All measured by Empatica E4 wristband. | 57 (30 M, 27 F) | UX (general) | VR (Oculus Rift DK2) | **In comparison** to 2-D group, the physiological response in terms of HRV and EDA were better for the VR groups.<br>**SCR** results of the participants showed an increased level in comparison to baseline. Even in a low-jerk content the slope of the increase is higher than high-jerk VR.<br>**Comparing** high jerk to low jerk and even non-immersive to immersive experience, showed a significant difference regards to HR and IBI.<br>**Less demand** of cognitive load is essential to improve the user quality of experience. |
| 14 | 2022 [14] | EEG - BIOSEMI active two device- data from 2 channels (C3 and C4). | 7 M | Sense of ownership | VR (Meta Quest2) | **There** may be both cognitive-sensory elements and experiential-belief elements in the sense of ownership. |
| 15 | 2022 [15] | EEG, ECG (HR, HRV), EOG (eye blink and saccade rate) all measured by Open BCI bioamplifier device.<br>11 EEG electrodes (Fp1, Fpz, Fp2, F3, F4, FCz, C3, C4, O1, Oz and O2) – 4 electrodes for EOG – 1 electrode for ECG - All signals with 125 Hz sampling rate. | 8 (5 M, 3 F) | En-gagement / arousal from EEG<br>Cybersick ness / sense of presence from EOG | VR (HTC VIVE Pro Eye) | **Physiological** measures could be used for real-time gamer quality of experience assessment. |
| 16 | 2022 [16] | ECG (HRV) – OpenBAN device - single channel - by Biosignalsplux 2 company - with three electrodes on | 19 (14 M, 5 F) | Cybersick ness | Google's Daydrea m View | **Significant** differences in the frequency components of HRV in response to cybersickness stimuli.<br>**Cybersickness** is a cumulative effect and reflected in the HRV measurements.<br>**Within** the limits of the experiment, it is not possible to know whether HRV is overestimating cybersickness or VSR (a |
| | | the chest – 1000 Hz sampling rate. | | | | questionnaire) is underestimating it. **Normalized** low frequency power (LFNU) and low to high ratio (LF/HF) frequency analyses data of HRV were the most important indicator of cybersickness. |
| 17 | 2022 [17] | EEG - Emotiv POC+ headset – 14 channels (AF3, F7, F3, FC5, T7, P7, O1, O2, P8, T8, FC6, F4, F8, and AF4) - 256 Hz sampling rate. ECG and EDA data with E4 Empatica device. | 18 (15 M, 2 F and 1 N/A) | Intuitiveness of the interaction method with the VR / cognitive load / engagement | VR (HTC VIVE) | **The** authors noticed an effect of interaction methods on the gamma band activities in the brain. **Facial** expression-based interaction group depicted characteristically higher peaks between 1–3 Hz (lower delta band), 16–22 Hz (Beta band), and 26–30 Hz (lower gamma band) than the controller-based interaction group. But in the delta and beta band, no significant differences were observed at p < 0.10. Whereas in the gamma band, significant differences were observed in all three environments at p < 0.10. **Significantly** higher activity in the gamma band of facial expression-based interaction group might suggest an increased cognitive load experienced by the participants. **Increased** beta and gamma band activity in both interaction mediums might suggest that the participants were actively engaged in the task. **For** Happy environment, facial expression caused higher EDA than controllers but it was the opposite in other two environments (neutral and scary). **Based on** the EDA data, the overall results illustrate that the controller seemed to elicit a greater reaction in the majority of environments. |
| 18 | 2022 [18] | EDA (SCR) - E4 Empatica wrist band - 4 Hz sampling rate. | 53 | Emotional engagement | | **skin** conductance, has potential as a tool for measuring emotional engagement. **Comparing** to non-VR situation, overall, the average of SCR was higher in the VR group (but not significantly different). **Overall** evaluation of post-experiment valence and arousal cannot be predicted from SCR-related variables including Peak, end, peak–end, and average SCRs. |
| 19 | 2022 [19] | HR, BT and EDA - Empatica E4 Wristband. | 11 | Sense of spatial presence | VR (PlayStation VR headset) | **After** adding an odour to the VR, average EDA is decreased from 3.93 to 2.22. **Adding** an odour (as a new cue) to a VR experience can significantly increase participants' sense of spatial presence. **Addition** of odour to a VR environment had a significant effect on both the psychological and physiological experience showing the addition of smell enhanced the VR environment. |
| 20 | 2022 [20] | EDA (GSR) - Empatica E4 wristwatch. | 11 (9 M, 2 F) | Emotional response | VR (HTC Vive Pro-Eye) | **The** authors reported a rise in the rate of emotional arousal as measured by the GSR sensor during the 3D learning simulation. **GSR** signal may be useful for investigating the causes of emotional arousal. |
| 21 | 2021 [21] | EEG (constructing multiband functional brain network efficiencies) - 64-channel – 1000 Hz sampling rate - 59 electrodes into five regions: frontal (F), parietal (P), occipital (O), left temporal (LT), and right temporal (RT). | 40 | UX (general) | VR | **At the** macro-scale, higher global efficiency in beta (16–32 Hz) and gamma (32–63 Hz) networks in the experiencing the 3-D environment in comparison to 2-D one. **At the** micro-scale, higher occipital and parietal efficiencies in beta and gamma networks in the 3D group, and higher frontal efficiency in the alpha (8–16 Hz) network in the 2D group. **Compared** to traditional 2D videos, 3D videos can result in different visual experiences. 3D environments are more realistic and immersive; however, some negative experiences, such as dizziness, headaches, and nausea, have also been reported. |
| 22 | 2021 | EMG data by | 189 (76 M, | Sense of | VR | **EMG** signals reflecting contraction of the |
| | [22] | EMG 100C amplifier (BIOPAC system) - positive and negative EMG input signals were placed on the right wrist (the flexor carpi radialis muscle). EEG data by EEG100C amplifier (BIOPAC system) - positive and negative input signals were placed on the left and right sides of the forehead. | 113 F) | presence | (HTC Vive Pro-eye) | brachioradialis muscle of the participants' arm, which was used as a body participation index reflecting the VR scene experience. **EMG** eigenvalue was significantly increased compared to the baseline level during the scene experience, indicating that the scene experience enhanced the contractive activities of the arm muscles, reflecting an increase in body involvement. **EMG** frequency domain characteristics were significantly different between the different scene experiences, indicating different degrees of arm muscle activity and body involvement in the different scenes, which supports the results there were differences in the sense of presence in the different scenes. **Three** EEG indices were calculated: BBR = $\beta_{high}/\beta_{low}$, EI = $\beta/(\theta + \alpha)$, and TBR= $\theta/\beta$, representing alertness, engagement, and calmness respectively. there were significant differences in the EEG index before and during the experience. **VR** restorative scene experience activated the prefrontal lobe, which is conducive to cognitive recovery. |
| 23 | 2021 [23] | EDA - Pip biosensor device– 8 Hz sampling rate. HR – Fitbit Charge device - 1 Hz sampling rate. | 60 (43 M, 17 F) | UX (general) / psychological arousal | AR (Microsoft HoloLens) / VR (Oculus Rift DK2) / Tablet (Lenovo 10-inch) | **In terms** of physiological response, similarities were found between the AR and tablet groups. This finding is interesting as both groups experience multimedia content whilst remaining in the context of a real-world setting. **Overall,** the results presented for physiological measurements reveal that the AR and tablet groups experience a higher quality of experience. **Although** the EDA reveals an increase in physiological arousal for the VR group, this may be attributed to limitations associated with the technology. **Implicit** measurements in the form of HR and EDA were observed as an insightful measure of physiological and, what is perceived to be, psychological arousal. |
| 24 | 2021 [24] | fNIRS - fNIR 400 device (Biopac Systems) - cerebral oxygenation (relative concentrations of average oxygenation values (µmol/L)) - prefrontal cortex – 2 Hz sampling rate - 16-channel - F7, Fp1, Fp2, and F8 and Brodmann's areas 9, 10, 45, and 46. | 20 (10 M, 10 F) | Cognitive demand | AR (Microsoft HoloLens) | **The** error rate in the designed interface of the AR, affected cerebral oxygenation parameters. **The** target size in the interface did not affect the cerebral oxygenation. **The** two-way interactions between the target size and error rate were not statistically significant, meaning that the effects of the error rates on the cerebral oxygenation parameters were consistent across the different target sizes. **These** results indicate the importance of the well-designed AR interactions in terms of lower system latency and appropriate target size to reduce the cognitive demand of the end-user. |
| 25 | 2021 [25] | EDA – HRV - SCR (mean and standard deviation) and ST (mean and standard | 32 M | Coherence / immersion | VR (nVisor SX) | **There is** strong evidence that exposure to bad coherence causes heart rate to increase. **Results** about the effect of immersion on physiological data was not consistent. |

| | | | | | |
|---|---|---|---|---|---|
| | | deviation) all measured by ProComp Infiniti. | | | |
| 26 | 2021 [26] | Event-related skin conductance responses (ER-SCR) - E4 Empatica wristband – 4 Hz sampling rate. | 15 (11 M, 4 F) | Transfer smoothness (Jerk of trajectory) | VR (HTC Vive) | **Jerk** affects user experience as its a crucial variable in the VR sickness. **Authors** noticed that the number of ER-SCR was smaller (p < .01) when the jerk equalled 2 m/s3 than when it equalled 1m /s3 and 3m /s3. |
| 27 | 2021 [27] | EMG from face muscles – assessing facial expression – **BioNomadix device** (Biopac) – **Root-mean-square (RMS) was used to quantify phasic fEMG responses – zygomaticus major muscle to measure smiling behavior and positively valanced emotion – 2 electrodes were used – 500 Hz sampling rate.** | 150 (72 M, 78 F) | Enjoyment | VR (Not declared ) | **Facial** EMG is an appropriate way to assess UX (especially enjoyment) of VR. **In comparison** to traditional video content, VR significantly enhances learning enjoyment (measured by EMG). |
| 28 | 2021 [28] | EEG – single channel (FP1) - NeuroSky Mind Wave device – 512 sampling rate. | 24 (7 M, 17 F) | Engageme nt | VR (Elemen ts DNA) | **Theta** band can be considered as a biomarker of the engagement in the VR environment. **EI score** (beta/(alpha+theta)) in the current study were not sensitive to VR engagement situation. |
| 29 | 2021 [29] | EEG - 21-channel device (BE micro)- 10 electrodes were used (Fpz, Fp1, Fp2, AFz, AF3, AF4, AF5, AF6, AF7, AF8) – 256 Hz sampling rate. | 42 (21 M, 21 F) | Mental effort | VR (HTC Vive) | **Theta** frequency band was the most informative band related to mental effort variations. **Results** showed a significant increment of mental effort for subjects exposed to the lavender scent in comparison to subjects exposed to the lemon scent during the entire virtual experience. |
| 30 | 2021 [30] | GSR (Equivital EQ02 from life monitor company) – had worn as a belt on the chest of the participants. | 16 M | Cybersick ness | VR (HTC Vive) | **For some** participants who suffered from higher sickness level motion sickness, had an increase in GSR amounts, however; this was not the case for some participants. **The increase** in GSR level for some participants was accompanied with an increase in motion sickness level, however; this correlation is not significant. |
| 31 | 2021 [31] | EEG/ERP (event-related potential) - ERP components | 37 (15 M, 22 F) | Sense pf presence | VR (Oculus GO) | **The** present study showed that late ERP components (slow waves = SW1 and SW2) recorded over the central brain may represent good electrophysiological correlates of the |

| | | | | | |
|---|---|---|---|---|---|
| | | evoked by the auditory stimuli - EEGO sport amplifier (ANT Neuro company) – 64-channel (Fz, Cz, POz electrodes used for sense of presence) – 512 Hz sampling rate - N1 (100–200 ms), MMN (150–200 ms), SW1 (400–650), and SW2 (650–900 ms) ERP components were analysed. | | | | subjective sense of presence. **The** low sense of presence participants' group (divided by 4 different questionnaire) showed a relative increase in all the components of interest (N1, MMN, SW1, and SW2). But in the later stage of analysis only slow waves remained significant. |
| 32 | 2021 [32] | EEG (Mynd Band BLE device) – 3 electrodes – 512 Hz sampling rate. | 31 | Sense of presence | VR (Oculus Rift) | **The** presence of the participants measured by EEG is much higher in VR than in PC. |
| 33 | 2021 [33] | Gaze (integrated Tobii eye tracking system into VR). HR (Scosche Rhythm armband monitor). | 18 (9 M, 9 F) | Cybersick ness | VR (HTC Vive Pro Eye) | **In this study**, the authors implied that, HR can be a suitable way to measure cybersickness. **Heatmap** of the gaze behaviour of the user indicated that, participants mostly concentrate to the mid part of the scene (±10 degree). So, removing peripheral parts of the virtual environment is logical. **In this** study, angular speed of the eyes of the participant was a good indicator of cybersickness. **The** result shows that, when you bluer thar peripheral part of the virtual scene, the amount of angular speed of the eyes decreased significantly. |
| 34 | 2020 [34] | EEG (by MUSE headset Edition 1). | 24 (first study, 18 M, 6 F) +16 (second study, 11 M, 5 F) | Engageme nt | VR (Oculus Rift CV1) | **In both** studies, EEG was a good indicator for engagement in virtual environment. **The** engagement index was $\beta/(\theta + \alpha)$ = averaged values of a, b, and h waves from the EEG Device. **In the** first study, authors implied that 12 different gestures, do not have any significant impact on engagement index measured by EEG. |
| 35 | 2020 [35] | Facial EMG – on the zygomatic major (i.e., the smiling muscle) for assessing joy - on the medial frontalis muscle for the measurement of fear. EDA (SCR) – 10k Hz sampling rate. | 63 (39 M, 21 F) | Affective response (Joy / fear by EMG Arousal by EDA) | VR (HTC Vive) | **Immersive** gameplay modes evoke a higher average amplitude of fear than less immersive gameplay modes. **Arousal** which is measured by EDA can be considered as a UX subdimension but it was not yielded consistent results. **Fear** elements within different gameplay modes evoked the experience of fear, arousal, and joy. **Feasibility** of leveraging on the temporal resolution of the psychophysiological methods to identify precise experience was proven. |
| 36 | 2020 [36] | EEG - Muse™ device – 4 electrodes consisting FP1 and FP2 (frontal lobe), TP9 and TP10 (temporal lobe) – 220 Hz sampling rate- Three EEG indexes is used separately including | 8 (3 M, 5 F) | Cybersick ness | Virtual environ ment by five 42-inch displays | **For the** following frequency bands and channels of EEG, the means of GF decreased significantly in motion sickness state: theta@FP1, alpha@TP9, alpha@FP2, alpha@TP10, and beta@FP1. **For the** following frequency bands and channels of EEG, the standard deviations of GF, which indicates the dispersion of the brain signal, decreased significantly in motion sickness state: alpha@TP9, alpha@FP1, alpha@FP2, alpha@TP10, and alpha@(FP2-FP1). **There was** a significant reduction in the means |
| | | Kolmogorov complexity (KC), Power spectral entropy (PSE) and Gravity frequency (GF). | | | | of KC for TP9, FP1, and FP2 when motion sickness onset.<br>**The values** of Cohen's d for each of the statistically significant t-tests were large (no less than 0.80), indicating that the EEG markers are indeed strong indicators of motion sickness.<br>**The experiments** demonstrate that an EEG device with a small number of electrodes (only four) is feasible to perform motion sickness evaluation. |
| 37 | 2020 [37] | EDA and HRV both measured by E4 wristband device – HR sampling rate = 1 Hz and EDA - 4 Hz sampling rate. | 33 (18 M, 15 F) | UX (general) | VR (Oculus Rift DK2) | **Low** jerk simulator, yielded in more positive EDA responses and lower heart rate (in comparison to high jerk simulation). |
| 38 | 2020 [38] | EEG (by epox flex from emotive company) – 32-channel device. | 9 | Sense of oneness for controlling human-machine system | VR (HTC vive) | **The** information flow in the brain were significantly reduced with the proposed visual haptics for the whole α, β, and θ-waves by 45% (in comparison to situation without visual haptics).<br>**The** result suggests that superimposing visual effects may be able to reduce the cognitive burden on the operator during the manipulation for the remote machine system. |
| 39 | 2020 [39] | Eye fixation – fixation count, fixation length, and time to fixation were calculated.<br>EDA – Number of peaks - Shimmer3 GSR+ device. | 72 (41 M, 31F) | Arousal | VR (HTC Vive Pro Eye) | **Eye** fixation data and number of peaks in GSR, can potentially used as biomarkers for studying virtual environments especially arousal of users who encountered with different virtual situations. |
| 40 | 2020 [40] | EEG by Enobio 32 device from Neuroelectrics company – 8 electrodes (FPz, F3, F4, Fz, P3, P4, Pz, Oz). | 15 (10 M, 5 F) | Cognitive demand | VR (HTC Vive) | **EEG** is a viable tool for future studies of presence and immersion in VR.<br>**EEG** spectral results show a trend towards the increased band power in all the examined bands (theta, alpha, upper beta), with the significance of the beta band, showing increased cognitive processing.<br>**Band power** in the frontal theta EEG band was significantly higher during the virtual experience (comparison to the baseline).<br>**Increase** in the beta band were identified as markers of the subjectively engaging VR experience. |
| 41 | 2020 [41] | EEG (by ENCEPHALAN-EEGR-19/26 device) – 26-channel – 19 brain regions (Fp1, Fp2, F7, F3, Fz, F4, F8, T3, C3, Cz, C4, T4, T5, P3, Pz, P4, T6, O1, O2) – 250 Hz sampling rate. | 5 M | Sense of presence | VR (HTC Vive) | **The** results showed the realism of a virtual reality simulator in the context of reproduction of the subjects' similar cognitive and semantic connections and motor programs.<br>**All** the cases there was a high level of activation in the frontal lobe, which can be explained by sensory and cognitive processing of information.<br>**There** was no brain activity in the range of 24−35 Hz (γ-waves), it was probably conditioned by high level skills of the ski instructors, and the experimental tasks did not require any intellectual efforts from them.<br>**There** was a difference determined between power spectrums obtained for the high-level of presence of about 10 Hz. At the same time, there was a difference determined between obtained power spectrums for low-level of presence more than 20 Hz. |
| 42 | 2020 [42] | EEG (by MOBITA – | 2 (1 M, 1 F) | Sense of agency | VR (passive | **The** results show the frontal and parietal brain regions as target areas during 3PP, mainly in β. |
| | | BIOPAC company) – 32-channel – 1000 Hz sampling rate – Only some regions (related to motor imagery) selected for analysis (FP1, FP2, F3, F4, Fz, C4, C3, Cz, C4, C3, Pz, T5, T6, O1, Oz, and O2). | | (controlling NAO robot) | HMD completed with galaxy S6 Samsung smartphone) | The node betweenness centrality shows us an important functional connection at C3 (primary motor cortex) in β, which could mean that C3 plays the role of regulator in the flow of information during the MI in 1PP condition. In α, the efficiency is greater in 1PP condition than in 3PP at the prefrontal cortex, which could be considered as a structural candidate for modulating inter-individual differences in the experience of presence. The results seem to indicate that α and β brain rhythms have a high indegree at prefrontal cortex in 1PP condition, and this could be possibly related to the experience of sense of agency. |
| 43 | 2019 [43] | EEG (B-Alert X10 device) – 256 Hz sampling rate – 9 channels were studied (Fz, F3, F4, Cz, C3, C4, Pz, P3, and P4). | 40 (14 M, 26 F) | Sense of presence | VR (HTC Vive) | The present study found that the feeling of presence during an affective VR experience using 360° video is associated with a decrease in EEG frontal alpha power and that this change is significantly correlated with the level of self-reported spatial presence. Results indicate that objective measures of EEG in frontal brain areas are reflective of the sense of presence, highlighting the potential of brain imaging techniques in assessing a user's experience in VR. |
| 44 | 2019 [44] | HR (Continuous RR) - intervals a pulse oximeter. | 80 | UX (general) | VR (not declared) | The configuration of fundamental interaction events 'Grab', 'Touch' and 'Move' has significant impact on VR UX. The usually concerned goal-based interaction event 'Operation' has secondly influence on VR UX. Meanwhile, it should be designed with dynamic difficulty adjustment for different user group, in order to increase the average positive impact of VR UX. |
| 45 | 2019 [45] | HR - Scosche Rhythm armband EDA – Mind field eSense Skin response sensor. | 20 M | HR for engagement and emotional state / EDA for emotional state | VR (Oculus Rift) | The fact of having already tried VR has a mild influence on participants' skin conductance level (higher value for novice users). Difficulty of a task can cause slight increase in HR which is the sign of involvement of the user. |
| 46 | 2019 [46] | ECG, Skin temperature and EDA all measured by BIOPAC's MP150 multi-channel – 1000 Hz sampling rate. | 30 (18 M, 12 F) | Emotional response | VR (not declared) | Volunteers in the VR environment have more obvious skin electrical signals. The experiment reveals the specific differences of user's emotional experience between the virtual reality and the traditional 2D environment from both subjective and objective aspects. The effectiveness of using the three physiological signals to measure emotions in the virtual environment is proven. |
| 47 | 2018 [47] | EEG – BIOSEMI Active Two device – 64-channel – 13 brain regions (Fp, AF, F, FT, FC, T, C, TP, CP, P, PO, O, I) – 2048 Hz sampling rate – 10 region of interests were selected based on used electrodes (FL = | 33 (22 M, 11 F) | Sense of presence | VR (HTC Vive) | The authors showed that the average time the brain needs to perceive and be adapted in a VE is 42.8 sec (SD=4.1 sec). This reveals that any application, developed in VR, needs to last at least for 42.8 seconds in order to be effective. The authors verified the role of parietal lobe in the VR experience, as an indicator of the sense of presence. The authors declared that our environment didn't require any direct interaction of the participant, we managed to notice the sense of presence, which means that interaction doesn't always play a significant role in the presence and the immersion. Of great interest is the presence of the theta state in the frontal region in both VR groups, |

| | | | | | |
|---|---|---|---|---|---|
| | | Frontal left, FC = Frontal center, FR = Frontal right, CL = Central left, CC = Central center, CR = Central right, PL = Parietal left, PC = Parietal center, PR = Parietal right, O = Occipital). | | | indicating the cognitive effort and representing the attentional processing (not shown before). **It signifies** that through a VR environment, educational processes might be more effective, as the cognitive efforts seems to be bigger. **The** dominance of the alpha band in the occipital lobe shows main visual attention mechanisms. |
| 48 | 2018 [48] | ECG (HRV measures), SCR and ST all measured by Pro Comp Infiniti device. | 32 (24 M, 8 F) + 32 M (for controlling confounding variable on the HRV regards to gender) | Immersion / coherence | VR (nVisor SX) | **The authors** implied that PI and Psi although previously were form an immersion experience, should consider different from each other. PI is the sign of immersion and Psi is the sign of coherence. **For** creating high and low immersion group of participants, the authors changed the field of view of the device from 60 to 30 degree respectively (changing PI). **The** Coherence factor was manipulated by changing the physical behavior of the environment (changing Psi). **It seemed** that the immersion manipulation (60° FoV vs. 30° FoV) did not have a noticeable effect on participants. **For** experiment 2, multiple system characteristics were varied simultaneously to create more sharply distinct high and low levels of immersion and coherence (not only FoV). **Skin conductance** as generally been considered to be less suitable as a measure of stress in virtual environment due to its slow onset and slow decay. **As** heart rate is affected by both stressful and "confusing" situations, whereas skin conductance seems only to respond to stress. |
| 49 | 2018 [49] | HR and EDA were measured with BITalino (r)evolution device. | 27 (10 M, 17 F) | Sense of presence | VR (Oculus Rift CV1) | **The** results showed that users had a lower HR mean variation in the dynamic stimulus condition. **Regarding** the EDA, the participants also had a lower EDA mean variation in the dynamic stimulus condition. **The** overall results of this study suggest that it is possible to modulate the sense of presence by linking changes in the VE to changes in the physiological response of the user and thus create physically sensitive VR experiences. |
| 50 | 2018 [50] | EEG – Live amp mobile amplifier – 30-channel – 500 Hz sampling rate. | 38 (18 M, 20 F) | Emotional arousal | VR (HTC Vive) | **The** results show that it is possible to predict subjective emotional arousal during a VR experience from brain activity. **An increase** in emotional arousal was associated with desynchronization of alpha oscillations in temporo-parietal areas. **Oscillatory** power in the high alpha and low beta range, particularly in temporo-parietal areas, might be the most promising marker. |
| 51 | 2018 [51] | EEG (only alpha wave) – Muse device (besides brain activity, it is capable to acquire muscle activity analysis of eyes and face). | 40 (20 M, 20 F) | Free navigation | VR (Oculus Rift DK2) | **The mean** alpha band power value for the multitouch screen was -0.024 Bels. This value was lower than the mean a value for the gamepad, which was -0.014 Bels. **The participants** blinked 0.26 times per second while running the navigation tasks using the multitouch screen. This blink rate was slower than the participants' blink rate while conducting the navigation tasks using the gamepad (0.314 times per second). |

| 52 | 2018 [52] | EEG (focused mainly on the frontal and parietal areas)- EMOTIV EPOC+ device – 14-channel. HR- fitbit charge 2 wrist band. | 25 (7 M, 18 F) | Sense of presence | VR (Oculus Rift CV) | **Alpha** state in parietal lobe was associated with sense of presence in the virtual class. **Increasing** in the HR of the participants, potentially can be a sign of sense of presence. |
|---|---|---|---|---|---|---|
| 53 | 2018 [53] | EEG (ERP=P300) - Biosemi Active Two system – 64-channel – 2048 Hz sampling rate - CPz, Pz and Poz regions received more attention than others. ECG - beats per minute. | 18 (15 M, 3 F) | Flow experience | VR (HTC Vive) | **The authors** found a decreased P300 and more high-frequency brain oscillations in VR compared to regular gaming (indicating more attention to the game). **Psychophysiological** measures are promising tools to quantify attentional allocation in VR, but more research is needed to clarify whether and how this translates to flow. **The results** suggests that the difference between regular and VR gaming in P300 wave is driven by the flow experience. **The difference** between PC and VR gaming was not significant for oscillatory power in the theta, alpha, low-beta and mid-beta frequency ranges. **There was** a significant increase for high-beta and gamma power spectral density in VR gaming compared to regular 2D gaming. **There was** no difference in average heart rate when participants were playing in VR compared to playing on the PC. |
| 54 | 2017 [54] | EEG - functional connectivity network in the theta and alpha frequency bands were calculated - EMOTIV EPOC system - 14 electrodes were attached (AF3, F7, F3, FC5, T7, P7, O1, O2, P8, T8, FC6, F4, F8, AF4) – 128 Hz sampling rate. | 50 M | Navigational skill | VR (not declared) | **During** acquisition of the spatial information, good navigators were distinguished by a lower degree of dispersion in the functional connectivity compared to the bad navigators (more functional connectivity). **People** with bad navigational skills showed a negative correlation between the recall score and percent connectivity during navigation. **Both** groups showed decrease in the functional connectivity during retrieval phase, demonstrating lesser recruitment of the resources as supported by fewer hubs activation. |
| 55 | 2017 [55] | EEG – functional brain connectivity was assessed - E-motive device– 128 Hz sampling rate – 16 electrodes. | 32 (10 M, 22 F) | UX (general) | VR (HTC vive) | **The frequency** between 40Hz and 50Hz has a significant difference between a game with good usability and a game with weak usability. **There were** significant differences mainly in the Gamma band regards to usability of the game (The synchronization levels for Gamma band were significantly higher during playing high usability game=UX has a different brain activity performance in Gamma frequency band). **Two games** were not different regards to Delta, Beat, Theta, Alpha1 and Alpha2 waves. **In relation** to the high and low usability conditions, the high usability condition showed an increase in frontal-parietal, frontal-temporal, and temporal-praetor-occipital PLI. Local frontal, central and temporal synchronizations were also higher in the high usability conditions. |
| 56 | 2016 [56] | ECG, GSR, and EMG were measured by BioPac MP150 device – 1000 Hz sampling | 31 M | Sense of presence | VR (Non-immersive 27-inch display) | **The analysis** of the psychophysiological data revealed that the first error message, the game crash, the control loss, and the drop to 4 fps resulted in a significant quadratic trend in the cardiac response curve. **The analysis** of the trapezius muscle as an |

| | | rate. | | | | indicator of startle responses indicated that only the interrupt break in oresence resulted in a statistically significant reaction.<br>**Activity** of the corrugator supercilii muscle was observed as an indicator of negative emotional processes. |
|---|---|---|---|---|---|---|
| 57 | 2015 [57] | HR- bipolar EL 504 Cloth Base electrodes from the left and right chest - Biopac MP150 1000 Hz sampling rate. Respiratory rate measured by Piezo-electric respiratory belt transducer. EMG - Zygomaticus major muscle from right part of the face . | 36 (20 M, 16 F) | Flow experienc e | VR (Not declared ) | **The results** revealed that HR, inter beat interval, HRV, low-frequency HRV (LF-HRV), high-frequency HRV (HF-HRV), and respiratory rate are all effective indicators in predicting flow experience. |
| 58 | 2015 [58] | EEG - EMOTIV EPOC – 14 electrodes | 15 (10 M, 5 F) | Emotional response | VR (*Oculus Rift*) | **Only** 4 users out of 15 felt a high excitement level in the course of the whole session, but 6 out of 15 declared they perceived a high emotional level. |
| 59 | 2015 [59] | EEG – 64-channel – 58 electrodes were recorded (from 58 electrodes, the data of 19 electrodes were analyzed: FP1, FP2, Fz, F3, F4, F7, F8, Cz, C3, C4, T3L, T4L, Pz, P3, P4, T5, T6, O1, O2) – 1000 Hz sampling rate. EMG from the real hand (extensor carpi radialis ECR) and flexor carpi ulnaris (FCU) muscles of both arms) - | 16 (10 M, 6 F) | Sense of agency (SoA) in using a virtual hand in the virtual environme nt | VR (with a non-immersi ve 21-inch monitor) | **Alpha band** activity may be the main neural oscillation of SoA, which suggests that the neural network within the anterior frontal area may be important in the generation of SoA. |
| 60 | 2014 [60] | EEG - Emotiv EPOC device – 14 electrodes were used. | 20 (11 M, 9 F) | Sense of presence | VR (a monitor and a wall) | **The Insula** region activation is related to stimulus attention and self-awareness processes, directly related with the sense of presence.<br>**Despite** the fact that the sense of presence questionnaire could not detect the difference between the two modes of monitor and wall in the level of presence perception, EEG was able to show this variable well. These results showed that to evaluate the user experience, objective methods should be used along with mental methods. |
| 61 | 2013 [61] | fMRI (1.5 T Siemens). | 14 F | Sense of presence | VR (MRI-compati ble video goggles) | **Frontal**, parietal and occipital regions showed their involvement during free virtual navigation.<br>**Activation** in the dorsolateral prefrontal cortex was also shown to be negatively correlated to sense of presence and the postcentral parietal cortex and insula showed a parametric increased activation according to the condition-related sense of presence, which suggests that stimulus attention and self-awareness processes related to the insula may be linked to the sense of presence. |

| 62 | 2013 [62] | EEG – 14-channel emotive EPOC. | 9 (6 M, 3 F) | Sense of presence | VR (non-immersive desktop screen) | **The results** showed significant differences between the free and automatic navigation conditions in the activity of right insula for the theta and alpha band. **The insula** activation is related to stimulus attention and self-awareness processes, directly related with the sense of presence. |
|----|-----------|--------------------------------|--------------|-------------------|-----------------------------------|---|
| 63 | 2013 [63] | EEG + NIRS. | 27 | Spatial free navigation | VR (virtual maze) | **NIRS** is an adequate method to measure brain correlates of spatial navigation in VR. increased fronto-central EEG theta oscillations (4-7 Hz) during spatial navigation in VR was found. **The results** of the NIRS measurement indicated increased deoxy-hemoglobin activation at parietal regions during VR navigation. **Oxy-Hb** over parietal areas was significant higher in bad performers compared to good performers. **EEG** theta power was negatively correlated with oxy- and deoxy-Hb over parietal areas. |
| 64 | 2009 [64] | Transcranial Doppler (TCD) sonography (for measuring brain activity) - 2 MHz pulsed-wave TCD unit – 100 Hz sampling rate. | 32 (24 M, 8 F) | Sense of presence | VR (cave-like walls) | **There were** changes in blood flow velocity in the subjects during moments associated with different levels of presence. |
| 65 | 2009 [65] | fMRI - 3T Siemens scanner. | 5 M | Immersion | VR (MRI-compatible screen) | **The results** indicated increased activity in both auditory and visual sensory cortices during multimodal (video plus audio) presentation. **Multimodal** presentation elicited increased activity in the hippocampus, a region well known to be involved in learning and memory. **Brain regions** associated with memory encoding would show increased activation in the more immersive multimodal condition. |
| 66 | 2002 [66] | HR – 3 electrodes SCR ST | 10 (3 M, 7 F) | Sense of presence | VR | **Change** in HR satisfied our requirements for a measure of presence, change in skin conductance did to a lesser extent, and that change in skin temperature did not. |

**M: male, F: female, BVP: Blood volume pressure, IBI: inter beat interval, HR: Heart rate, ResR: respiration rate, SCR: skin conductance response, GSR: galvanic skin response, EDA: electro dermal activity, ST: skin temperature, BT: body temperature, VR: virtual reality, AR: augmented reality, fMRI: functional magnetic resonance imaging, EEG: electroencephalography, EOG: electrooculography, EMG: electromyography, fNIRS: functional near infrared stereoscopy.**

As it is clear from the model proposed in this study, the UX of virtual environments has several subsets, including positive, negative and neutral experiences. In the table above, the physiological parameters related to each UX are listed separately. Below, the information is presented in summary.

**Sense of Presence:** The feeling of being present in the virtual environment is the shortest description of this experience. As mentioned above, various methods have been proposed to measure it. In general, however, EEG has been the most widely used method to quantify it, measuring electrical activity from the surface of the skull. As with other mental processing activities, the (pre)frontal cortex is involved. One of the studies referred to AF7 and AF8 regions in this regard. In addition, another study related cognitive retrieval theories to the effectiveness of Virtual Reality-based therapies, as retrieval occurs in the frontal cortex. Other regions of interest were the insula (increased activity) and the postcentral parietal cortex (increased activity). In addition, activation of the dorsolateral prefrontal cortex implies a decreased sense of presence in the virtual environment. From another view and based on frequency analysis, alpha, theta/beta, beta/(theta+alpha), Beta $_{high}$/ Beta $_{low}$, theta (in the prefrontal region), decrease alpha (in frontal), increase alpha (in parietal) were considered biomarkers of the sense of presence.

**Immersion:** Previously, we said that sometimes immersion is used interchangeably with the sense of presence. But it is wrong to say that immersion is one device-dependent feature and another human-dependent feature. In this case, the authors in [79] supported this claim and modified the physical characteristics of the device to change people's immersion. A variety of physiological biomarkers/devices have been used to quantify this immersion, including EDA, HRV, EEG, fMRI, and ST. With this variety of methods, an adequate objective method to quantify UX has not been documented in most cases. Only in the study with the fMRI device was it shown that the brain region related to memory encoding, e.g., the hippocampus, was related to immersion. Although EEG was successful in quantifying other user experiences, the use of this device failed to introduce good biomarkers.

**Engagement:** It is equivalent to being actively involved in a 3D experience. A review of the results in the table above shows that 3 devices or indicators EEG, EDA and HR were used to quantify it. In the two studies that used heart rate, one considered it an adequate indicator, i.e., increased heart rate was a sign of involvement, while the other found no significant relationship between it and involvement. Two studies on EDA were the same as the previous index. One study showed that phasic EDA data could be considered an adequate indicator, that is, an increase in EDA was a sign of engagement, while another study found no significant difference. But as with most indicators, EEG was the most widely used device. In one study, the authors introduced increased beta and gamma waves as indicators. In another study, only theta waves were reported for this concept. The most challenging indicator was beta / (alpha+theta) which was accurately used as an indicator in one study and denied in another.

**Cognitive load:** As in all interactions humans have with other technologies, in the interaction of humans with virtual environments the mental pressure on users should not be excessive. In fact, users should not consume too many cognitive resources while using these products. Therefore, mental pressure is classified as one of the important dimensions of UX. The table above shows that 3 studies have addressed this issue through EEG devices. The most important indicators related to mental demand were gamma frequency (increase), theta (in the frontal region),alpha (generally through the scalp) and upper beta (generally through the scalp). Another study used an fNIRS device and implicated the role of hemoglobin assessment as a biomarker of mental demand.

**Flow experience:** It is compromised by the positive and effortless experiences of a virtual environment. Although one study showed that this UX cannot be measured through heart rate, two other articles revealed the possibility of using heart-related parameters to quantify flow. As shown in the table above, the two indicators of LF and SDNN related to end-users' HRV can be considered biomarkers of this UX. In addition, another study demonstrated the role of the p300 wave (central part of the parietal lobe) of EEG-ERP as an important biomarker of flow experience.

**Sense of Agency:** This is the sensation of controlling an object in the virtual environment. As shown in the table above, three studies were devoted to this experience and all used EEG. Based on the spectral analysis of frequencies, alphas (increase in the whole brain, especially in the anterior frontal part), gamma (increase in the whole brain) and beta (in the frontal and parietal regions) were the most important biomarkers of this use experience. In addition, because the primary motor cortex region of the brain (C3) plays a crucial role in this experience, one study implicated the role of beta frequency in this region.

**Cybersickness:** Because of the nature of interaction with virtual spaces, cybersickness is a major component of the UX of these products, although it represents a negative and annoying feeling. It can be said that, just like the "sense of presence," the negative experience of cybersickness has also been well described by researchers. Several physiological indicators and devices have been introduced to measure this UX, including EEG, EOG and ECG. Regarding EOG, one study reported the suitability of eyelid blink and saccade velocity, while another study reported eye angular velocity as three main physiological indicators of cybersickness. HR and HRV (LFNU and LF/HF indices) were also discussed in two separate studies. One research reported the ability of GSR to quantify cybersickness. A direct positive relationship was found between these two items. This means that an increase in GSR is a sign of user cybersickness. As shown in the table above, in [67], the electrical activity of the brain, particularly the prefrontal and temporal areas, is considered an important biomarker of cybersickness.

**Free navigation:** In this case, the end UXs a free path within a virtual environment. EEG alone or in combination with the fNIRS device has been the possible solution to quantify this experience. In one study, brain alpha power was introduced as a suitable biomarker. Another study showed increased increased theta (in the fronto-central region) as a biomarker. The increase in the amount of deoxyhemoglobin in the parietal region was also introduced as a reliable indicator of this UX.

**Coherence:** Modification of the physical behavior of the virtual environment primarily results in plausibility that is related to the coherence of an interaction. Although too many physiological parameters have been tested to quantify this experience, only HR, increased in a noncoherent scenario, can be considered a reliable biomarker.

So far, a summary has been presented for UXs that were mentioned more than once in the table above. For some of them, however, which are mentioned only once or are general concepts, such as arousal and emotional response, no summary is provided. In any case, the table provides sufficient information for all user experiences according to the model.

## 5. Discussion

### 5.1 UX and not Usability

Our review of the literature from the past 20 years (2002-2023) allows us to make a state of the art regarding biomarkers and psychophysiological indicators substantial for quantifying UX within virtual environments.

When it comes to UX, readers unconsciously mix a concept such as usability. One of the strengths of the present study was the attempt to separate these two concepts, which are two close but different concepts [98, 99]. User eXperience, as reflected in its "user" component, indicates a characteristic related to a user in using a product or service. But just as clearly, the main components of usability, while having some subsets related to user opinion, include features related to the scope of the product or service. Therefore, the most widely accepted definition of usability, presented in the form of the ISO 9241 standard, obtains information about the efficiency and effectiveness of the system and finally measures user satisfaction. This decision was useful for this study from another point of view, because it avoided the analysis of a huge number of unnecessary studies reporting only the efficiency and effectiveness of VR products.

## 5.2 UX concept in virtual environments

User eXperience in VR environments has been the subject of attention for several years. Of course, we have already mentioned that despite the introduction of questionnaire measurement tools or the presentation of several models, none of them has covered the concept of UX in virtual environments comprehensively or nearly comprehensively.

Tcha-Tokey et al. [100] proposed a holistic model of UX in virtual environments. In this model, only a few subdimensions of the UX in the virtual environment were grouped together, including presence, immersion, engagement, flow, and emotion. As is evident, some crucial user experiences were omitted in this model, which were addressed in the present study. In addition, other researchers have tried to use famous technology-related models, such as the Technology Acceptance Model (TAM), to include users' interaction with virtual environments [101]. But even in this approach, some undeniable subdimensions of UX are ignored. In a more detailed view, some of these user experiences have been modeled independently or, finally, with one or two other components that, although valid, are not completely comprehensive [102, 103].

## 5.3 UX sub-dimensions for virtual environments (covered by questionnaires)

For some components of UX in the proposed model, there are already questionnaires for their measurement. Among them, we can refer to questionnaires for measuring cybersickness/motion sickness in VR [104], embodiment [105], immersion [106], etc. Questionnaires have also been created for some complications caused by the nature of VR use [107].

In addition, to study the general concept of UX in these environments, questionnaires such as the standard UEQ, also in short or modular versions, and "Attrakdiff 2" have been used, which were also previously developed for the general concept of UX and which, not being specific to virtual environments, do not have perfect dimensions.

## 5.4 Strength of proposed model (QoUX in VE)

The conceptual model presented in this study attempted to specify a considerable number (26 items) of components of UX in a VR environment. Most of these subsets were also well evaluated with users' physiological parameters, which are also categorized in the results section.

## 5.5 Limitation

The model presented in this study is conceptual, and although some cause-and-effect relationships among its subsets are clear or even conceivable, this model should be fully tested in empirical and field studies, which were not among the objectives of this study.

# 6.Conclusion

In this study, we examined the potential of physiological metrics to quantify end-users' User eXperience of the virtual environment, as it became apparent that subjective evaluation and questionnaires do not allow for a comprehensive investigation. Easy access to physiological indices for each of the UX subdimensions can guide researchers to fill previous gaps or strengthen weak evidence.

**Acknowledgments:** Members of the Usability lab are very acknowledged for their responsive comments to improve the quality of this manuscript.

**declaration of interest statement:** Nothing declared.